\documentstyle[prl,aps,preprint,epsf]{revtex} 
\large  

\begin{document}
\draft

\title{Continuum theory of vacancy-mediated diffusion} 
\author{T. J. Newman} 
\address{Department of Physics,\\ 
Virginia Polytechnic Institute and State University,\\ 
Blacksburg VA 24061, USA} 
\maketitle
\begin{abstract}
We present and solve a continuum theory of vacancy-mediated diffusion
(as evidenced, for example, in the vacancy driven motion of tracers
in crystals). Results are obtained for all spatial dimensions, and reveal the
strongly non-gaussian nature of the tracer fluctuations. In integer
dimensions, our results are in complete agreement with those from previous
exact lattice calculations. We also extend our model to describe
the vacancy-driven fluctuations of a slaved flux line.
\end{abstract}
\vspace{5mm} 

\newpage

\section{Introduction}

Random walks and associated diffusion processes are
ubiquitous in Nature\cite{hug}. Our understanding of their
properties has grown steadily over the years, with 
contributions from workers in many fields, but there are always
surprises in store as we learn to ask new and more
probing questions. 

An interesting example of a non-trivial
random walk problem is that of vacancy-mediated diffusion (VMD)
\cite{vmdr,vmd}.
This applies to the case of a crystal containing a low
density of vacancies; the question being, what are the
fluctuations of a tagged particle (or impurity) which moves only due
to exchange with wandering vacancies? This process
has a more generic application as it is one of the simplest `slaved
diffusion processes'. It has been studied in
various guises over the years\cite{vmd} and a lattice formulation
was solved exactly by Brummelhuis and Hilhorst \cite{bh} 
in the late Eighties. The most revealing aspect of their solution is
the strongly {\it non-gaussian} nature of the tagged particle's fluctuations.
The same lattice model received 
alternative exact treatments recently \cite{toro}.

Given one is generally interested in the scaling behaviour
of such systems, it is useful to have a more coarse-grained
treatment to hand which allows one to access the 
(hopefully universal) long-wavelength behaviour directly,
without having to resort to a sophisticated lattice
calculation. A well-known example of such a situation is the critical 
behaviour of a ferromagnet\cite{fe}, which may be studied within the context
of the lattice (Ising) model (where an exact treatment
is only possible in dimensions $d=1,2$), or at a more 
coarse-grained level via the $\phi ^{4}$ field theory (where
controlled calculations can estimate the exponents in the
most interesting case $d=3$)\cite{wk}.

The purpose of this paper is to construct and analyse a 
continuum (or coarse-grained) theory of vacancy-mediated
diffusion. Fortunately the continuum theory is exactly
solvable in all dimensions, which allows a direct comparison
with the previous lattice calculations\cite{bh} -- complete agreement
is found. Given this situation, one has confidence in
applying the continuum theory to situations where a lattice
formulation would be more difficult, if not intractable. We
consider one such situation here; namely the motion of
a directed (flux) line under the action of diffusing vacancies.

The outline of the paper is as follows. In the next section
we define the lattice model of vacancy-mediated diffusion
more precisely, and, using this as a base, construct the
continuum theory. We then highlight the `mesoscopic' physics
contained within this coarse-grained model. In Section III
we analyse the continuum model at the level of mean-field theory (MFT).
The apparent physics within MFT is simple, and the solutions
of the mean-field equation yield results which contain some
correct scaling information (i.e. the length-time scaling),
but completely miss the more interesting statistical aspects; 
namely the strongly non-gaussian nature of the fluctuations
as found from the lattice calculations. Thus we are led to 
attempt an exact solution of the continuum theory. This is
made possible by formulating an infinite-order perturbation
theory, as described in Section IV. The results for various
dimensions are derived and presented in Section V -- the
non-gaussian statistics are seen to be completely reproduced,
with the added advantage that the results from the continuum 
theory are valid for arbitrary dimension $d$. In Section VI
we give an illustration of the utility of this coarse-grained
approach. We formulate a continuum model for 
the transverse fluctuations of a $d+1$ directed (flux) line 
due to VMD of pinning centres.
We end the paper with Section VII which contains
our conclusions, along with some suggested extensions of the 
present work.

\section{Formulation of the continuum theory}

We first consider a simple lattice formulation of VMD
with which we shall motivate our continuum theory.
To be specific, let us consider the following lattice
model. Each site ${\bf r}$ of a $d$-dimensional hypercubic lattice
contains a spin $S_{\bf r}$ which may take the values $\pm 1$.
[In the crystal lattice application one would take all spins as
`up', thus referring to crystal atoms, except for the tagged
particle (or impurity) which would be assigned a `down' spin.] 
A spin may only alter its value when involved in an
exchange with a single diffusing vacancy located at
the position ${\bf R}(t)$. As only the {\it position} of the
vacancy is relevant we may give the vacancy a spin $+1$
for notational convenience. [We shall assume for simplicity
that only one vacancy is present -- the generalization
to many vacancies is straightforward.] There are various
ways to describe this model. For instance, one can define
a probability distribution for the vacancy position and
the position of the `up' and `down' spins, and then write
a master equation for its evolution. Alternatively, one
can regard the process as a stochastic cellular automata (SCA),
and attempt to write explicit SCA rules for its operation.
We shall follow the latter approach. 

The rule for updating the vacancy position is written as
\begin{equation}
\label{dvupdate}
{\bf R}(t+\delta t) = {\bf R}(t) + {\bf l}(t) \ ,
\end{equation}
where ${\bf l}(t)$ is a unit lattice vector drawn with
equal probability from the $2d$ possible choices.
The update of the spins is written down as follows: a
spin will remain unchanged unless it is located either 
at the vacancy position at time $t$, or at the
subsequent vacancy position at time $t+\delta t$.
Thus we have
\begin{equation}
\label{dsupdate1}
S_{\bf r}(t+\delta t) = S_{\bf r}(t) + 
\delta _{{\bf r},{\bf R}(t)}[S_{{\bf R}(t+\delta t)}(t)-S_{\bf r}(t)] + 
\delta _{{\bf r},{\bf R}(t+\delta t)}[S_{{\bf R}(t)}(t)-S_{\bf r}(t)] \ ,
\end{equation}
which may be re-arranged (with the help of (\ref{dvupdate}))
in the appealing form
\begin{equation}
\label{dsupdate2}
S_{\bf r}(t+\delta t) - S_{\bf r}(t) = 
[S_{{\bf R}(t)+{\bf l}(t)}(t)-S_{{\bf R}(t)}(t)] \  
[\delta _{{\bf r},{\bf R}(t)}-\delta _{{\bf r},{\bf R}(t)+{\bf l}(t)}] \ .
\end{equation}

The factorization of the SCA update for $S_{\bf r}(t)$ gives
us reason to hope that a simple Taylor expansion may be
invoked to yield an accurate continuum limit, since the
leading order terms will appear in a multiplicative
fashion (as opposed to terms occurring in an additive
fashion, in which case one is unsure of their relative
importance). Thus we shall take the simplest continuum limit:
First, the position of the vacancy is described by a
real vector ${\bf R}(t) \in {\cal R}^{d}$ satisfying
\begin{equation}
\label{cvupdate}
d{\bf R}/dt = {\bf \xi}(t) \ ,
\end{equation}
which is the continuum equivalent of Eq.(\ref{dvupdate}).
The random lattice vector ${\bf l}$ has given way to a
random vector ${\bf \xi}$ (with an implicit factor of $\sqrt{\delta t}$)
which is drawn from a gaussian distribution $P[\xi]$ with zero mean, and 
covariance $\langle \xi ^{\alpha} (t)\xi ^{\beta}(t') \rangle =
D\delta _{\alpha,\beta}
\delta (t-t')$. [Henceforth, angle brackets denote an average over $P$.]
Second, the spin variable $S_{\bf r}(t)$ will be replaced by a coarse-grained
`magnetization density', or `order parameter' (OP) denoted by the
field $\phi ({\bf r},t)$. Expanding each term of Eq.(\ref{dsupdate2})
to first order, we obtain
\begin{equation}
\label{csupdate}
\partial _{t} \phi ({\bf r},t) = -\lambda \left [ {\bf \xi} \cdot
{\partial \phi ({\bf R}(t),t) \over \partial {\bf R}(t)} \right ] \ 
\partial _{t} \delta ^{d} ({\bf r} - {\bf R}(t)) \ ,
\end{equation}
where $\lambda $ is a phenomenological parameter with dimensions $L^{d}T$.
This last equation represents our continuum theory of VMD. It {\it loosely}
resembles a Langevin equation, but the extremely implicit appearances
of the noise ${\bf \xi}$ throughout the equation forbid such a simple
designation. For instance, it is not clear how one would write down
a dynamical equation for the OP probability distribution $P[\phi, t]$ --
it would certainly not fit within the standard Fokker-Planck category.
However, taxonomy is of little importance to us. We shall accept the
equation on its own merits. As a first step in this direction, let us
probe its more obvious physical content.

We can ascribe an independent meaning to each of the two factors on
the right-hand-side of Eq.(\ref{csupdate}). The second of the two
is simple enough in spirit -- it allows temporal change of
the OP only in the neighbourhood of the vacancy position ${\bf R(t)}$.
The first factor guarantees (via the directional derivative) 
that the OP temporally changes
only when the vacancy moves through a region in which the OP has spatial 
variation. Furthermore, the amount of temporal variation is linearly
coupled to the amount of spatial variation (with a strength $\lambda $).
This appears reasonable with regard to the physical properties of
VMD we wish to model. Whether one would have written 
down such a continuum model
{\it a priori} based on these considerations is unclear. However, given
the simple `derivation' from the SCA, one finds the {\it a posteriori}
physical motivations satisfactory.

It is useful to note that analytic
progress on Eq.(\ref{csupdate}) is very difficult without first
recasting it in Fourier space (so as to rid ourselves of the implicit
nature of the noise). 
Denoting the Fourier transform (FT) of the OP by ${\tilde \phi}({\bf k},t)$
we have
\begin{equation}
\label{csupdateft}
\partial _{t} {\tilde \phi} ({\bf k},t) = -\lambda \ {\tilde G}({\bf k},t)
\int dk_{1} \ {\tilde G}({\bf k}_{1},t)^{*}  {\tilde \phi} ({\bf k}_{1},t) \ ,
\end{equation}
where 
\begin{equation}
\label{gfn}
{\tilde G}({\bf k},t)=({\bf \xi} \cdot {\bf k}) 
\exp [i{\bf k}\cdot {\bf R}(t)] \ , 
\end{equation}
and $dk \equiv d^{d}k/(2\pi)^{d}$.

We shall now discuss the initial conditions. As for the vacancy, we simply
need to define its initial position ${\bf R}_{0} \equiv {\bf R}(t=0)$. The
initial condition $\phi _{0}({\bf r})$
for the OP will vary according to the physical system
we are interested in modelling. As regards the case of a tagged particle
(or impurity) being driven by the wandering vacancy, we take the
initial condition on the lattice to be all spins up, bar one down spin
(representing the tag) located at some point (the origin, say). In the 
continuum, this may
be described by 
\begin{equation}
\label{initc}
\phi _{0}({\bf r}) = A - B\delta ^{d}({\bf r}) \ .
\end{equation}
Alternative scenarios may be investigated by modifying the initial
OP distribution. For instance, setting $\phi _{0}$ to be
a step function would be appropriate for modelling the vacancy
mediated roughening of an initially straight domain wall\cite{sch}.

Once the initial conditions are defined we may formally integrate the
equations of motion to give
\begin{equation}
\label{cvupdate2}
{\bf R}(t) = {\bf R}_{0} + \int \limits _{0}^{t}dt' \ {\bf \xi}(t') 
\end{equation}
and
\begin{equation}
\label{csupdateft2}
{\tilde \phi} ({\bf k},t) = {\tilde \phi}_{0} ({\bf k}) 
-\lambda \int \limits _{0}^{t} dt' \ {\tilde G}({\bf k},t')
\int dk_{1} \ {\tilde G}({\bf k}_{1},t')^{*}  
{\tilde \phi} ({\bf k}_{1},t') \ .
\end{equation}
This completes our formulation of the model. The main focus of this
paper is to calculate the mean OP density 
$\rho ({\bf r},t) = \langle \phi ({\bf r},t) \rangle$,
which is the quantity analogous to the tagged particle distribution function
calculated in previous lattice studies. In the next section,
we shall present a brief mean-field analysis of $\rho $, whilst sections
IV and V contain a description of the exact solution for $\rho $ from
Eqs.(\ref{cvupdate2}) and (\ref{csupdateft2}) above.

\section{Mean field theory}

The purpose of this section is to indicate how much one can learn
about the system from a simple, yet uncontrolled, mean field theory (MFT).
We shall find that MFT predicts the correct length-time scaling,
but misses the non-gaussian nature of the fluctuations in VMD. This error
persists in all dimensions (bar $d=0$). 

We shall define the MFT to be used here in an operational sense. Namely,
we perform the simplest possible average of the OP equation of motion
(\ref{csupdateft}), by replacing the average of the right-hand-side
by the product of two separate averages. Explicitly, we have
\begin{equation}
\label{mft1}
\partial _{t} {\tilde \rho }({\bf k},t) = -\lambda \ \int dk_{1} \ 
\langle {\tilde G}({\bf k},t){\tilde G}({\bf k}_{1},t)^{*} \rangle \
{\tilde \rho} ({\bf k}_{1},t) \ .
\end{equation} 
We refer the reader to appendix A for the evaluation of 
$\langle {\tilde G}({\bf k},t){\tilde G}({\bf k}_{1},t)^{*} \rangle $. 
The averaging necessarily introduces a temporal cut-off $t_{0}$ (which sets
the implicit correlation scale of the white noise $\xi $). We shall only
ever work to leading order in $1/t_{0}$. Eq.(\ref{mft1}) now takes the form
\begin{equation}
\label{mft2}
\partial _{t} {\tilde \rho }({\bf k},t) = -{\lambda D\over t_{0}}
k^{\alpha} \ \int dk_{1} \ k^{\alpha}_{1} \
{\tilde \rho} ({\bf k}_{1},t) \ e^{-(D/2)({\bf k}-{\bf k}_{1})^{2}t + 
i({\bf k}-{\bf k}_{1})\cdot {\bf R}_{0}} \ ,
\end{equation} 
where the momentum components are indicated by a superscript for
future notational convenience.

This integral equation may be recast in a more illuminating form by
inverse Fourier transforming the density. After some re-arranging, one has
\begin{equation}
\label{mft3}
\partial _{t} \rho ({\bf r},t) = -{\lambda D\over t_{0}}
\int dk_{1} \ k^{\alpha }_{1} \ e^{-i{\bf k}_{1}\cdot {\bf r}}
{\tilde \rho} ({\bf k}_{1},t) \int dk \ (k^{\alpha}+k^{\alpha }_{1})
e^{-(D/2)k^{2}t - i{\bf k}\cdot ({\bf r}-{\bf R}_{0})} \ .
\end{equation} 
Now the inner integral may be written as 
$(i\partial _{r^{\alpha }} + k^{\alpha }_{1})g({\bf r}-{\bf R}_{0},t)$. 
The Green function 
$g({\bf r},t)=(2\pi Dt)^{-d/2} \exp [-r^{2}/2Dt]$ is the probability density
for the vacancy (in other words, $g$ is the solution of the
Fokker-Planck equation corresponding to Eq.(\ref{cvupdate})).
One may then manipulate the outer integral over ${\bf k}_{1}$ 
in terms of the inverse FT
of the density to obtain the following partial differential equation
for $\rho ({\bf r},t)$ :
\begin{equation}
\label{mft4}
\partial _{t} \rho = {\lambda D \over t_{0}}
\nabla \cdot \left [ g({\bf r}-{\bf R}_{0},t) \nabla \right ] \ \rho \ \ \  .
\end{equation}

As a MFT, the above equation makes good sense. It contains the simple
physical information that the OP density undergoes a diffusion process,
but with the twist that the diffusivity is not constant, but proportional
to the probability density of the wandering vacancy. One would have had 
little trouble in writing down such a MFT using physical arguments alone.

A complete analytic treatment of Eq.(\ref{mft4}) is beyond our
present remit. However, the limit $r^{2} \ll Dt$ is trivially
solved, since to leading order the equation reduces to
\begin{equation}
\label{mft5}
\epsilon (t) \partial _{t} \rho = {D\over 2} \nabla ^{2} \rho \ ,
\end{equation}
where $\epsilon (t) = (t_{0}/2 \lambda)(2\pi Dt)^{d/2}$.
One may solve the above equation using FT, and with the initial
condition specified in Eq.(\ref{initc}) we obtain 
$\rho = A-Bg({\bf r}, \tau (t))$, where $\tau (t)$ is the
effective time scale of the tagged particle, and is given
by $\tau (t) = \int \limits _{t_{0}}^{t} dt' \ \epsilon (t')^{-1} \ $
which, ignoring numerical prefactors, takes the form
\begin{equation}
\label{mft6}
\nonumber
\tau (t) \sim \left \{
\begin{array}{ll}
& {\lambda \over (Dt)^{d/2}} {t \over t_{0}} \ \ \ \ \ \ \ \ 0 < d < 2 \\
& {\lambda \over Dt_{0}} \ln (t/t_{0}) \ \ \ \ d = 2 \\
& {\lambda \over (Dt_{0})^{d/2}} \ \ \ \ \ \ \ \ \ \ d > 2
\end{array}
\right. 
\end{equation}

This solution firstly tells us that the OP density always has
a gaussian envelope (at least for $r^{2} \ll Dt$, which encompasses most
of the physically interesting scales, since $t \gg \tau$), 
but that the temporal spreading of the
envelope is not that of a random walker ($\Delta r \sim \sqrt t$), but 
much reduced. In two dimensions the spreading increases only logarithmically
in time (as the vacancy's random walk is only just recurrent), and
for $d>2$ the spreading halts altogether after some finite time, which
is accounted for by the vacancy having `fled the scene', never to return.

The main interest in VMD is not so much in the renormalized time scale,
which one can argue for on simple physical grounds, but rather in the
non-gaussian nature of the OP fluctuations. These non-gaussian fluctuations
were exactly calculated in lattice theories, and were generally found
to have tails which decay {\it slower} than a gaussian. It is the aim
of the following two sections to reproduce these features from the
continuum theory.

\section{Solution via perturbation theory}

We have seen the relative failure of MFT, which is essentially due
to imposing a strict locality on the OP equation of motion. In fact, 
the evolution of the OP is non-local as seen from Eq.(\ref{csupdate}).
A more systematic treatment is required to handle the subtle correlations
between vacancy and OP. Given the functional nature of Eq.(\ref{csupdate}), 
the most useful analytic technique would appear to be perturbation theory.
Since the equation of motion is linear, we shall not encounter an
exponentially divergent number of terms at higher orders; rather, each order
will contain only a single term. 

So referring to the time integrated (and Fourier transformed) evolution
equation for the OP, namely Eq.(\ref{csupdateft2}), we make the
substitution
\begin{equation}
\label{subpt}
{\tilde \phi}({\bf k},t) = \sum \limits _{n=0}^{\infty} \lambda ^{n}
{\tilde \chi }_{n}({\bf k},t) \ , 
\end{equation}
where ${\tilde \chi }_{0}({\bf k},t) = {\tilde \phi}_{0}({\bf k})$.
Equating powers of the coupling $\lambda $ yields (for $n>0$)
\begin{equation}
\label{ptrel}
{\tilde \chi}_{n}({\bf k},t) = -\int \limits _{0}^{t} dt' \ 
{\tilde G}({\bf k},t') \int dk' \ {\tilde G}({\bf k}',t')^{*}
{\tilde \chi }_{n-1}({\bf k}',t') \ .  
\end{equation}
This relation may be iterated to give the explicit solution for
each order of perturbation theory as
\begin{eqnarray}
\label{ptsol}
\nonumber
{\tilde \chi}_{n}({\bf k},t) = (-1)^{n} \int \limits _{0}^{t} dt_{1}
\cdots \int \limits _{0}^{t_{n-1}} dt_{n} \ {\tilde G}({\bf k},t_{1}) 
& & \left [ \prod \limits _{m=1}^{n-1} \int dk_{m} \ 
{\tilde G}({\bf k}_{m},t_{m})^{*}{\tilde G}({\bf k}_{m},t_{m+1}) \right ]\\
& & \ \ \ \ \ \times \int dk_{n} {\tilde G}({\bf k}_{n},t_{n})^{*}
{\tilde \phi }_{0}({\bf k}_{n}) \ .  
\end{eqnarray}
 
In principle, the solution given above may be used to calculate
a range of spatio-temporal OP correlation functions. Our present aim is
more modest -- we shall perform a direct average of each term in the
perturbation series in order to obtain the OP density $\rho ({\bf r},t)$.
We have
\begin{equation}
\label{ptden}
{\tilde \rho}({\bf k},t) = {\tilde \phi }_{0}({\bf k}) + 
\sum \limits _{n=1}^{\infty} \lambda ^{n}
\langle {\tilde \chi }_{n}({\bf k},t) \rangle \ . 
\end{equation}
Using the explicit form for the ${\tilde G}$ functions as given in
Eq.(\ref{gfn}) we may write 
\begin{equation}
\label{ptav}
\langle {\tilde \chi }_{n}({\bf k},t) \rangle = \int \limits _{0}^{t} dt_{1}
\cdots \int \limits _{0}^{t_{n-1}} dt_{n} \ k^{\alpha _{1}} \left [
\prod \limits _{m=1}^{n-1} \int dk_{m} \ 
k_{m}^{\beta _{m}}k_{m}^{\alpha _{m+1}} 
\right ] \int dk_{n} k_{n}^{\beta _{n}}{\tilde \phi}_{0}({\bf k}_{n}) \ 
Q_{n}(\lbrace {\bf k}_{m},t_{m};\alpha _{m},\beta _{m}\rbrace ) \ ,
\end{equation}
where the function $Q_{n}$ represents the following average: 
\begin{equation}
\label{qfn}
Q_{n}(\lbrace {\bf k}_{m},t_{m};\alpha _{m},\beta _{m}\rbrace ) = (-1)^{n}
\left \langle \prod \limits _{m=1}^{n} \xi ^{\alpha _{m}}\xi ^{\beta _{m}}
\exp \left [ i{\bf R}(t_{m}) \cdot 
({\bf k}_{m-1}-{\bf k}_{m}) \right ] \right \rangle \ ,
\end{equation}
with the time ordering $t_{1} \ge t_{2} \ge \cdots \ge t_{n}$, and the
notation ${\bf k}_{0}\equiv{\bf k}$.
At this stage of the calculation we can see clearly the remaining steps.
First, we must perform a multivariate average in order to determine
the function $Q_{n}$. Second, we must perform the $n$-fold integral
over the momenta $\lbrace {\bf k}_{m} \rbrace $. Third, we must perform
the $n$-fold integral over the intermediate times  $\lbrace t_{m} \rbrace $.
Finally, we are left to resum the functions ${\tilde \chi}_{n}$ which
will yield the FT of the mean OP density. In the absence of hindsight, it 
is somewhat surprising that all these steps may be performed exactly 
for arbitrary
dimension $d$. The remainder of this section will consist of the details
of the first two steps, whilst the third and fourth steps will be presented
in the subsequent section as they are dimension-specific. Henceforth, we
shall take the initial position of the vacancy to be at the origin: 
${\bf R}_{0}={\bf 0}$. This slight loss of generality (which is of no
physical significance in the long-time regime) is more than compensated
by calculational simplicity.

We begin with the first step; that of determining $Q_{n}$. We refer the 
reader to Appendix B in which this multivariate average is evaluated.
The result is
\begin{eqnarray}
\label{qfn2}
\nonumber
Q_{n}(\lbrace {\bf k}_{m},t_{m};\alpha _{m},\beta _{m}\rbrace ) 
= \left (- {D \over t_{0}} \right ) ^{n} \prod \limits _{m=1}^{n} 
\Biggl \lbrace \Bigl [ \delta _{\alpha _{m},\beta _{m}} 
& - & Dt_{0}(k^{\alpha _{m}}-k_{m}^{\alpha _{m}})
(k^{\beta _{m}}-k_{m}^{\beta _{m}}) \Bigr ] \\
& & \times\exp \Bigl [- {D\over 2}({\bf k}-{\bf k}_{m})^{2}
(t_{m}-t_{m+1}) \Bigr ] \Biggr \rbrace \ ,
\end{eqnarray}
where the symbol $t_{n+1} \equiv 0$.

To proceed with the momentum integrals, it is necessary to decide upon
an initial condition. We shall use that given in Eq.(\ref{initc}), since
our interest is presently focussed on the vacancy-driven diffusion of
a tagged particle.
The generic momentum integral (for $1 \le m \le n-1$) takes the form:
\begin{eqnarray}
\label{momint1}
\nonumber
\int & dk_{m} & \ k_{m}^{\beta _{m}}k_{m}^{\alpha _{m+1}} 
\Bigl [ \delta _{\alpha _{m},\beta _{m}} 
- Dt_{0}(k^{\alpha _{m}}-k_{m}^{\alpha _{m}})
(k^{\beta _{m}}-k_{m}^{\beta _{m}}) \Bigr ]
\exp \Bigl [- {D\over 2}({\bf k}-{\bf k}_{m})^{2}
(t_{m}-t_{m+1}) \Bigr ] \\
& = & [D(t_{m}-t_{m+1})]^{-1}[2\pi D(t_{m}-t_{m+1})]^{-d/2} 
\Bigl [\delta _{\alpha _{m},\alpha _{m+1}} + D(t_{m}-t_{m+1})
k^{\alpha _{m}}k^{\alpha _{m+1}} \Bigr ] \ ,
\end{eqnarray}
where we have worked to leading order in $(1/t_{0})$.
The final momentum integral involves the OP initial condition which
in Fourier space takes the form
${\tilde \phi}_{0}({\bf k}) = (2\pi )^{d} A \delta ^{d}({\bf k}) - B$.
Its evaluation yields
\begin{eqnarray}
\label{momint2}
\nonumber
\int dk_{n} \ k_{n}^{\beta _{n}}{\tilde \phi}_{0}({\bf k}_{n}) 
\Bigl [ \delta _{\alpha _{n},\beta _{n}} 
- Dt_{0}(k^{\alpha _{n}}-k_{n}^{\alpha _{n}})
(k^{\beta _{n}}-k_{n}^{\beta _{n}}) \Bigr ] 
\exp \Bigl [&-&{D\over 2}({\bf k}-{\bf k}_{n})^{2}t_{n} \Bigr ]  
\ \ \ \ \ \ \ \ \ \ \ \ \ \ \ \ \ \ \\
& = & \ \ \ -Bk^{\alpha _{n}}(2\pi Dt_{n})^{-d/2} \ ,
\end{eqnarray}
to leading order in $(1/t_{0})$.
Collecting our results from Eqs.(\ref{qfn2}), (\ref{momint1}) and 
(\ref{momint2}), and substituting back into (\ref{ptav}), we have
\begin{eqnarray}
\label{chifin}
\nonumber
\langle {\tilde \chi }_{n}({\bf k},t) \rangle = 
-BD[-(2\pi D)^{\gamma }t_{0}]^{-n} & & 
\int \limits _{0}^{t} dt_{1} \Biggl [ \int \limits _{0}^{t_{1}} 
dt_{2} (t_{1}-t_{2})^{-(1+\gamma )}
\cdots \int \limits _{0}^{t_{n-1}} dt_{n} (t_{n-1}-t_{n})^{-(1+\gamma )}
\Biggr ] t_{n}^{-\gamma } \\
& & \times k^{\alpha _{1}} \Biggl \lbrace \prod \limits _{m=1}^{n-1} 
[\delta _{\alpha _{m},\alpha _{m+1}} + D(t_{m}-t_{m+1})k^{\alpha _{m}}
k^{\alpha _{m+1}} ] \Biggr \rbrace k^{\alpha _{n}} \ ,
\end{eqnarray}
where we have introduced $\gamma \equiv d/2$. An important point
must be mentioned at this stage. The intermediate time integrals
above appear to be divergent at one or both of their lower and upper
limits. This divergence is regularized by the existence
of the microscopic time scale $t_{0}$. As mentioned before, this scale
appears as an effective correlation time in the white noise process. It
may be taken to be arbitrarily small (with respect to any `experimental' time
scale in which one is interested). Therefore, any time integral limit
is naturally softened by $t_{0}$.

We have now completed two of the four steps in the calculation of
${\tilde \rho }({\bf k},t)$. To proceed further requires the
evaluation of the $n$-fold integral over the intermediate times
which is sensitive to spatial dimension, and is presented in the
next section.

\newpage

\section{Mean OP density in various dimensions}

As mentioned above, we must now specify the spatial dimension of
interest. In fact there are three cases: i) $d>2$, ii) $d=2$, and
iii) $d<2$. These cases were already apparent within MFT, and arise
because of the qualitatively different behaviour of the vacancy's
random walk. In case i), the vacancy will essentially `disappear' from
the vicinity of the tagged particle after a finite time. In case ii),
the vacancy's walk is marginally recurrent and we expect a slow, but steady,
evolution of the OP density for all times. In case iii),
the vacancy's walk is `strongly' recurrent, and thus the cross-section
of vacancy-tag collisions is always `large'. It is the aim of this final
stage of the calculation to replace these qualitative descriptions
by precise results. We shall analyse the three cases in turn.

\subsection{$d > 2$}

In this and the following two subsections we shall take advantage of the
smallness of $t_{0}$. Integrals which are apparently divergent will be
regularized using $t_{0}$, and only the most singular contribution will
be retained. We stress that this form of regularization is not a
mathematical manoeuvre, but is entirely consistent with the physical
meaning of white noise; namely a noise process which is a limiting
form of a microscopic process with a correlation time $t_{0}$. 

Referring to Eq.(\ref{chifin}) we adopt the following strategy to
evaluate the integrals. First, we explicitly contract the $n$-fold
momentum product, which will yield $2^{n-1}$ terms in $n$ sets; terms in
the $m^{\rm th}$ set being characterized by a factor of $k^{2(m+1)}$
(where $m$ counts from 0 to $n-1$). A 
term in the $m^{\rm th}$ set will also carry a string composed of 
$m$ different factors of the form $(t_{j}-t_{j+1})$. For
each term, we perform the time integrals in order, starting from $t_{n}$, 
keeping only the most singular contribution at each step, and being careful to
include the appropriate time-difference factors in the numerator 
(from the string).
This procedure is fairly simple for $d>2$, since the only integrals
one encounters are
\begin{eqnarray}
\label{intdgt2}
\nonumber
\int \limits _{t_{0}}^{t-t_{0}} ds \ {1\over (t-s)^{1+\gamma }s^{\gamma }}
& \simeq & {1\over \gamma }{1\over (t_{0}t)^{\gamma }} \ , \\
\int \limits _{t_{0}}^{t-t_{0}} ds \ {1\over (t-s)^{\gamma }s^{\gamma }}
& \simeq & {2\over (\gamma -1)}{1\over t_{0}^{\gamma -1}t^{\gamma }} \ , \\
\nonumber 
\int \limits _{t_{0}}^{t-t_{0}} ds \ {1\over s^{\gamma }}
& \simeq & {1\over (\gamma -1)}{1\over t_{0}^{\gamma -1}} \ .
\end{eqnarray}

One finds that each term of a given set has the same value after integration.
In detail, the $m^{\rm th}$ set contains $C^{n-1}_{m}$ equal terms of value
$$D^{m}k^{2(m+1)}\left ( {1\over \gamma t_{0}^{\gamma }} \right )^{n-m-1}
\left ( {2\over (\gamma -1)t_{0}^{\gamma -1} } \right )^{m}
\left ( {1\over (\gamma -1)t_{0}^{\gamma -1}} \right ) \ . $$
Thus, the series composed of the $n$ sets is nothing more than a 
binomial series; which is trivially summed. The dominant contribution
from the $n$-fold time integral may therefore be combined with the 
constant prefactor of Eq.(\ref{chifin}) to give
\begin{equation}
\label{chifin2}
\langle {\tilde \chi }_{n}({\bf k},t) \rangle = 
{BDk^{2} \over (\gamma -1)(2\pi Dt_{0})^{\gamma }}
\left \lbrace - {1 \over \gamma t_{0}(2\pi Dt_{0})^{\gamma }}
\left [ 1 + {2\gamma \over (\gamma -1)}Dk^{2}t_{0} \right ] 
\right \rbrace ^{n-1} \ .
\end{equation}

This ends the third step (namely the intermediate time integrals). The
last step is to reconstruct ${\tilde \rho}$ from the infinite sum
given in Eq.(\ref{ptden}), using Eq.(\ref{chifin2}) above.
In the present case, this sum is seen to be simple, as the series 
(in powers of $\lambda $) is geometric. 
Explicitly evaluating the sum yields the final result in the form
\begin{equation}
\label{denfindgt2}
{\tilde \rho }({\bf k},t) = A (2\pi )^{d} \delta ^{d}({\bf k}) 
- {B \over 2} \left [ 1 + {1\over 1 + (k/\Lambda )^{2}} \right ] \ ,
\end{equation}
where the momentum scale is given by
\begin{equation}
\label{momscaledgt2}
\Lambda ^{2} = {(\gamma -1)\over 2Dt_{0}} \left [ 1 + {(2\pi Dt_{0})^{\gamma }
t_{0} \over \lambda } \right ] \ .
\end{equation}
It is of interest to recast this result in real space. For instance,
in the physically pertinent case of $d=3$ one has
\begin{equation}
\label{dendgt2}
\rho ({\bf r},t) = A  
- {B \over 2} \left [ \delta ^{3}({\bf r}) + {\Lambda ^{2} \over 4\pi r}
e^{-\Lambda r} \right ] \ .
\end{equation}

Thus, for $d>2$ we find that the initial  $\delta $-function of the OP
density is evolved such that only half of its weight is smeared. 
That half which is smeared attains a Lorentzian profile in Fourier space, 
with a momentum scale $\Lambda $ as given in Eq.(\ref{momscaledgt2}). This
scale is seen to be an effective UV cutoff ($1/\sqrt {Dt_{0}}$) renormalized
by the vacancy-tag coupling $\lambda $. Note that this scale is independent
of time, in accordance with our expectations. Note also that the smearing of
the OP density is strongly non-gaussian. In real space, the smearing creates
an OP density which is exponential in form, as seen explicitly for $d=3$
above. 

One final point: we expect that as $d$ increases, less of the 
initial $\delta $-function (in the OP density) will be smeared, since the
vacancy will disappear from its vicinity with increasing efficiency. The
fact that we find exactly half of the $\delta $-function to be smeared,
for all $d$, is a consequence of starting the vacancy's walk precisely
at the location of the $\delta $-function. Had we chosen ${\bf R}_{0}
\ne {\bf 0}$, the $d$-dependence of the `smearing fraction' would have been
apparent.

\subsection{$d = 2$}

We now turn to the marginal case of $d=2$. Within MFT we found that
the root-mean-square fluctuations of the tag (i.e. the smearing of
the initial $\delta $-function in the OP density) grow as $[\ln (t)]^{1/2}$.
We expect this slow growth to be retained within the exact solution.
Our main interest is in how the functional form of the OP density
differs from the gaussian found in MFT.

In exactly two dimensions, the $n^{\rm th}$-order contribution to the OP
density, as given in Eq.(\ref{chifin}), takes the form
\begin{eqnarray}
\label{chifin2d}
\nonumber
\langle {\tilde \chi }_{n}({\bf k},t) \rangle = 
-BD(-2\pi Dt_{0})^{-n} & & 
\int \limits _{0}^{t} dt_{1} \Biggl [ \int \limits _{0}^{t_{1}} 
dt_{2} (t_{1}-t_{2})^{-2}
\cdots \int \limits _{0}^{t_{n-1}} dt_{n} (t_{n-1}-t_{n})^{-2}
\Biggr ] t_{n}^{-1 } \\
& & \times k^{\alpha _{1}} \Biggl \lbrace \prod \limits _{m=1}^{n-1} 
[\delta _{\alpha _{m},\alpha _{m+1}} + D(t_{m}-t_{m+1})k^{\alpha _{m}}
k^{\alpha _{m+1}} ] \Biggr \rbrace k^{\alpha _{n}} \ .
\end{eqnarray}
Our strategy for evaluating the time integrals is the same as before.
We multiply out the integrand to form a total of $2^{n-1}$ terms
arranged in $n$ sets. Each term is integrated over the $n$
intermediate times, with only the most singular piece retained from
each integral. Care is taken to include the appropriate factors, for
a given term, in the numerator. The integrals one encounters are given
below (with $p \ge 0$):
\begin{eqnarray}
\label{int2d}
\nonumber
\int \limits _{t_{0}}^{t-t_{0}} ds \ {[\ln (s/t_{0})]^{p} \over (t-s)^{2}s }
& \simeq & {[\ln (t/t_{0})]^{p} \over t_{0}t} \ , \\
\int \limits _{t_{0}}^{t-t_{0}} ds \ {[\ln (s/t_{0})]^{p} \over (t-s)s}
& \simeq & {(p+2)\over (p+1)} {[\ln (t/t_{0})]^{p+1}\over t} \ , \\
\nonumber 
\int \limits _{t_{0}}^{t-t_{0}} ds \ {[\ln (s/t_{0})]^{p} \over s}
& \simeq & {1\over (p+1)} [\ln (t/t_{0})]^{p+1} \ .
\end{eqnarray}

As before, one finds that each term of a given set has the
same value after integration. In this case the $m^{\rm th}$ set 
contains $C^{n-1}_{m}$ equal terms of value
$$D^{m}k^{2(m+1)}(1/t_{0})^{n-m-1}[\ln (t/t_{0})]^{m+1} \ .$$ 
The $n$ sets form a binomial series which is trivially summed. We find
\begin{equation}
\label{chifinwe}
\langle {\tilde \chi }_{n}({\bf k},t) \rangle = 
{BDk^{2}\ln (t/t_{0}) \over 2\pi Dt_{0}}
\left \lbrace - {1 \over 2\pi Dt_{0}^{2}}
\left [ 1 + Dk^{2}t_{0}\ln (t/t_{0}) \right ] 
\right \rbrace ^{n-1} \ .
\end{equation}
The final step is to sum over the functions 
$\langle {\tilde \chi}_{n} \rangle$ as
prescribed by Eq.(\ref{ptden}). As before, this series is geometric
and the sum may be immediately performed to give
\begin{equation}
\label{denfin2d}
{\tilde \rho }({\bf k},t) = A (2\pi )^{d} \delta ^{d}({\bf k}) 
- B \left [ {1\over 1 + (k/\Lambda )^{2}\ln (t/t_{0})} \right ] \ ,
\end{equation}
where the momentum scale is given by
\begin{equation}
\label{momscale2d}
\Lambda ^{2} = {1 \over Dt_{0}} \left [ 1 + {2\pi Dt_{0}^{2}
\over \lambda } \right ] \ .
\end{equation}
Inverting the FT yields our final result
\begin{equation}
\label{den2d}
\rho ({\bf r},t) = A - {B\Lambda ^{2} \over 2\pi \ln (t/t_{0})}
{\rm K}_{0} \left [ {\Lambda r \over [\ln (t/t_{0})]^{1/2}} \right ] \ ,
\end{equation}
where ${\rm K}_{0}$ is the modified Bessel function of zeroth order 
\cite{as}.

We see that the gaussian envelope for the spreading of the OP density
(as found in MFT), has given way to a completely different form, namely
the Bessel function ${\rm K}_{0}$. 
This is in complete agreement with the previous exact lattice calculations
\cite{bh}.

\subsection{$d < 2$}

Finally we consider VMD for $d<2$. Within lattice calculations the
case of $d=1$ may be studied either within the geometry of a chain, or 
within the geometry 
of a strip of finite width. In the former case, the situation is
trivial, since the tag can only be moved back and forth to one of two 
sites as the vacancy passes by. In the latter, the smearing of the
tag distribution function is non-trivial; the gaussian fluctuations of
MFT giving way to a stretched exponential. In this subsection, we shall 
see how to recover these results, along with their generalization to
arbitrary $d \in [0,2]$.

Our starting point is the expression for ${\tilde \chi}_{n}$ given
in Eq.(\ref{chifin}). We shall adopt the same strategy as before
to evaluate the $n$-fold integral over the intermediate times.
In this case we encounter the following integrals (with $p \ge 0$)
\begin{eqnarray}
\label{intdlt2}
\nonumber
\int \limits _{t_{0}}^{t-t_{0}} ds \ {1\over (t-s)^{1+\gamma }
s^{(p+1)\gamma - p}}
& \simeq & {1\over \gamma t_{0}^{\gamma }t^{(p+1)\gamma - p}} \ , \\
\int \limits _{t_{0}}^{t-t_{0}} ds \ {1\over (t-s)^{\gamma }
s^{(p+1)\gamma - p}}
& \simeq & {B(1-\gamma,(p+1)(1-\gamma ))
\over t^{(p+2)\gamma -(p+1)}} \ , \\
\nonumber 
\int \limits _{t_{0}}^{t-t_{0}} ds \ {1\over s^{(p+1)\gamma - p}}
& \simeq & {t^{(p+1)(1-\gamma )}\over (p+1)(1-\gamma )} \ ,
\end{eqnarray}
where $B(a,b)$ is the Beta function\cite{as}.

In the two previous subsections, we were able to extract the most singular
contributions from the $2^{n-1}$ terms in the integrand of Eq.(\ref{chifin}),
and we found that these contributions formed a binomial series which was
then easily summed. In the present case, this simple summability is lost,
due to the presence of the Beta functions. However, we retain the feature
that each of the $C^{n-1}_{m}$ terms within the $m^{\rm th}$ set are
equal in value. Thus, after some manipulations, we may reduce the 
function $\langle {\tilde \chi }_{n} \rangle $ to the form
\begin{equation}
\label{chidlt2}
\langle {\tilde \chi }_{n}({\bf k},t) \rangle = 
-{B \over [-\gamma t_{0}(2\pi Dt_{0})^{\gamma }]^{n}}
\sum \limits _{m=0}^{n-1} C^{n-1}_{p} {[\gamma \Gamma (1-\gamma )Dk^{2}
t^{1-\gamma}t_{0}^{\gamma }]^{m+1} \over \Gamma (1+(m+1)(1-\gamma ))} \ .
\end{equation}
In order to perform the summation over $m$, we introduce Hankel's
representation of the Gamma function\cite{as}:
\begin{equation}
\label{hank}
{1\over \Gamma (z)} = {i\over 2\pi} \int _{C} d\tau \ e^{\tau } \tau ^{-z} \ ,
\end{equation}
where the contour $C$ runs from minus infinity above the negative real axis,
encircles the origin clockwise, and then returns to minus infinity below
the real axis. Using this representation for the Gamma function appearing
in the denominator of the summand of Eq.(\ref{chidlt2}), we may explicitly 
perform the (binomial) sum. Each function $\langle {\tilde \chi }_{n}\rangle $ 
now takes the form of
an integral over $\tau $, with $n$ appearing only as a simple power
in the integrand. Thus the sum over these functions 
(as dictated by Eq.(\ref{ptden})) is again geometric and may be performed 
with ease. One then has the following integral expression for the
FT of the mean OP density, valid for $0 < d < 2$ ({\it i.e.}
$0 < \gamma < 1$):
\begin{equation}
\label{ftdendlt2}
{\tilde \rho }({\bf k},t) = A(2\pi )^{d}\delta ^{d}({\bf k})
-B{i \over 2\pi} \int _{C} d\tau \ {e^{\tau} \over \tau ^{\gamma }}
\left [ {1 \over \tau ^{1-\gamma } + (k/\Lambda )^{2}(t/t_{0})^{1-\gamma }}
\right ] \ ,
\end{equation}
where the renormalized UV cutoff is given by
\begin{equation}
\label{uvdlt2}
\Lambda ^{2} = {1\over \gamma \Gamma (1-\gamma ) Dt_{0}}
\left [ 1 + {\gamma (2\pi Dt_{0})^\gamma t_{0} \over \lambda }\right ] \ .
\end{equation}

We now wish to extract the scaling behaviour of the mean OP density from
the above expression. For convenience we define 
$\delta \rho = A-\rho $, which is initially a $\delta$-function with
amplitude $B$. Let us first specialize to $d=1$. In this
case one may simplify the above integral considerably, using a procedure
outlined in Appendix C. The result is the following scaling function:
\begin{equation}
\label{scalfn1d}
\delta \rho ({\bf r},t) = {B \over \pi} \ {z \over r} \ 
\int \limits _{0}^{\infty } {ds \over \sqrt{s}} \exp [ -s^{2} - z^{2}/4s]
\ ,
\end{equation}
where the scaling variable is $z=r\Lambda /(4t/t_{0})^{1/4}$.
This integral is easily analyzed for both $z \ll 1$ and $z \gg 1$.
In the former case, one finds
\begin{equation}
\label{scalsmz}
\delta \rho ({\bf r},t) = {B \over 2\sqrt{2} \ \pi } \Lambda \left (
{t_{0}\over t} \right )^{1/4} \left [ \Gamma (1/4) - 2\sqrt{\pi} \ z 
+ O(z^{2}) \right ] \ , 
\end{equation}
whilst in the latter, a steepest descents analysis yields
\begin{equation}
\label{scallgz}
\delta \rho ({\bf r},t) \sim {B z^{2/3}\over \sqrt{\pi} \ r} 
\exp [-(3/4)z^{4/3}] \ . 
\end{equation}
Both of the above results are in complete agreement with the scaling
functions found by Brummelhuis and Hilhorst\cite{bh}, from an exact lattice
calculation for VMD on an infinite strip. This gives us 
strong confidence in the physical integrity of our continuum theory of VMD.

For completeness we briefly describe the form of the mean OP density
for arbitrary dimension $d \in [0,2]$. The scaling variable in this
case is generalized to 
\begin{equation}
\label{scalvar}
z(\gamma ) = \left ( {1-\gamma \over 2} \right )^{(1-\gamma)/2}
{r\Lambda \over (t/t_{0})^{(1-\gamma )/2}} \ . 
\end{equation}
Referring to Eq.(\ref{ftdendlt2}), inverse FT and subsequent analysis
yields the following results.
For $z(\gamma ) \ll 1$ we find
\begin{equation}
\label{scaldsmz}
\delta \rho ({\bf r},t) = {B \over (4\pi )^{\gamma }} {\Gamma (1-\gamma)
\over \Gamma (\gamma + (1-\gamma)^{2})} \Lambda ^{2\gamma } \left (
t_{0} \over t \right )^{\gamma (1-\gamma )}\left [ 1 + O(z^{2(1-\gamma )})
\right ]  \ , 
\end{equation}
whilst a steepest descents analysis for $z(\gamma) \gg 1$ reveals
\begin{equation}
\label{scaldlgz}
\delta \rho ({\bf r},t) \sim \exp \left [-{(1+\gamma )\over 2}
z(\gamma )^{2/(1+\gamma )} \right ] \ .
\end{equation}

This completes our study of the simplest VMD scenario; namely the
effective diffusion of a tagged particle by a wandering vacancy.
We have been able to give exact results for all dimensions. In
the case of integer dimensions, our results are found to be in
complete agreement with previous exact lattice studies.

\section{Extension to a slaved flux-line}

In this penultimate section we shall consider a more complicated
VMD scenario, both as an illustration of the utility
of the coarse-grained approach, and also as a physical model
for a VMD mechanism of flux lattice melting.

We consider an anisotropic lattice consisting of well separated planes
(in particular, one of the high-$T_{c}$ cuprates\cite{cup}).
Within each plane there exists a low density of wandering vacancies.
We now imagine a flux line directed perpendicular to the planes, and
strongly pinned by certain lattice impurities\cite{pin}. If the binding energy
is strong, thermal wandering of the line will be completely suppressed.
However, a much weaker form of line wandering may be driven by VMD
of the pinning sites themselves (due to exchange with the low density
planar vacancies, or vacancy aggregates). 
Since each plane has its own stock of vacancies,
(which we disallow from hopping from plane to plane), the interactions between
a given vacancy and the line are recurrent and we can expect slow and
steady smearing of the mean density of the flux line. The physical
existence and/or relevance of this mechanism deserves more
detailed investigation (for instance, there are several competing
pinning mechanisms within the material, one of which is actually due
to the oxygen vacancies themselves\cite{pin}).

We shall describe this system by generalizing the continuum theory
of VMD outlined in section II. First, we take for simplicity one
vacancy within each plane. In the continuum, this is simply described
by attaching a longitudinal coordinate to the vacancy position ${\bf R}$.
The equation of motion for the vacancies is then given by
\begin{equation}
\label{vaceq}
\partial _{t}{\bf R}(z,t) = {\bf \xi }(z,t) \ ,
\end{equation}
where the noise has zero mean and covariance $\langle \xi ^{\alpha}(z,t)
\xi ^{\beta}(z',t') \rangle = 
D\delta _{\alpha,\beta}\delta (z-z')\delta (t-t')$.
The OP $\phi $ now describes the probability
density of the flux line. It is a function of a planar coordinate
${\bf r}$, a longitudinal coordinate $z$, and time $t$. For a given
$z$, the evolution of the OP is given by Eq.(\ref{csupdate}), in two
dimensions. The simplest longitudinal coupling shall be taken -- namely,
an elastic interaction (which stems from the Josephson coupling
between Cu-O planes\cite{ld}). Thus the equation of motion for the OP is
\begin{equation}
\label{csupdatefl}
\partial _{t} \phi ({\bf r},z,t) = \nu \partial^{2}_{z}\phi ({\bf r},z,t) 
-\lambda \left [ {\bf \xi} \cdot
{\partial \phi ({\bf R}(z,t),z,t) \over \partial {\bf R}(z,t)} \right ] \ 
\partial _{t} \delta ^{d} ({\bf r} - {\bf R}(z,t)) \ ,
\end{equation}
where $\nu $ is an effective longitudinal elasticity.

As an initial condition, the simplest choice is to take the
flux line to be straight, and located at the origin: $\phi ({\bf r},z,0) = 
A\delta ({\bf r})$. Also, we could (artificially) start all the vacancies at 
the origin: ${\bf R}(z,0)={\bf 0}$.

This model may now be analysed in precisely the same way as our original
VMD model; namely through an infinite order perturbation expansion in powers 
of $\lambda$. One must use an additional longitudinal FT to diagonalize the
elastic coupling. The appearance of multiple longitudinal Green functions
makes the explicit evaluation of the functions 
$\langle {\tilde \chi}_{n} \rangle $ more challenging than before. However,
analytic progress seems possible and is currently being pursued. It is
certainly of interest to calculate the $\nu $-dependence of the 
mean OP evolution, to see how effectively the flux line elasticity
combats the driving forces of VMD. Preliminary results indicate that
i) for $\nu \rightarrow \infty $ the line fluctuations become gaussian
and coincide exactly
with the fluctuations of a single point within MFT (as described in
section III), and ii) for $\nu \rightarrow 0$ the line fluctuations
have a {\it singular} dependence on the elasticity (and thus differ
from the fluctuations of independent planar tags as described in 
sections IV and V).

\section{Conclusions}

In this paper we have constructed and solved a continuum theory of
vacancy-mediated diffusion. Our main intention has been to test
the theory against exact results known from lattice studies\cite{bh,toro}.
In particular we have thoroughly examined the evolution of the mean
OP density in the case of a single vacancy smearing an originally
sharply-peaked OP fluctuation (which corresponds to the lattice
scenario of following the motion of a tagged particle due to vacancy
exchange). At the level of mean field theory, we found the (almost
obvious) length-time scaling for the OP density, but no sign of the
non-gaussian fluctuations, which were the most interesting results
obtained from the lattice studies. Therefore we pursued a 
more systematic treatment, based on an infinite order perturbation
expansion. We presented an exact analysis of our theory 
in all dimensions, and complete agreement has been found with the 
dynamical scaling results obtained from the lattice. In particular, we
find that for $d>2$, the OP density is smeared over a limited range and
then freezes. The envelope is a simple exponential. In $d=2$, the
smearing is slow, but continues indefinitely. The envelope is
described by the modified Bessel function ${\rm K}_{0}$. Finally, for
$d<2$, a more challenging calculation revealed that the envelope of
OP smearing is described by a stretched exponential. Our results have
the advantage of being valid for arbitrary dimension, thus revealing
more clearly their analytic structure. In the final section we 
proposed an extension of simple VMD to the diffusion of a pinned flux
line, slaved to planar vacancy exchange. Analysis of this more 
challenging problem is in progress.

There are, needless to say, many other extensions to the current work. 
We mention
the more obvious here. First, it would be interesting to calculate
higher order OP correlations. Throughout the present work, we have
concentrated on the mean OP density. However, there is much
non-trivial information hidden in the simplest spatio-temporal correlation
functions. Second, one can apply the continuum theory to more complicated
(single vacancy) scenarios; mainly by adjusting the boundary and
initial conditions. For instance, a step function initial condition would
correspond to the vacancy-mediated roughening of an initially straight 
domain wall. Finally, one could investigate simple mechanisms whereby 
the random walk of the vacancy itself is weakly coupled to the
OP distribution, which is a physically relevant perturbation.

\vspace{1.0cm}

The author would like to thank Z. Toroczkai and R. Zia for bringing
this problem to his attention. The author also thanks E. Lundell
for a critical reading of the manuscript.
The author gratefully acknowledges financial support from the
Division of Materials Research of the National Science Foundation.

\newpage

\appendix

\section{}
In this appendix we explicitly evaluate the average
$\langle {\tilde G}({\bf k},t){\tilde G}({\bf k}_{1},t)^{*} \rangle $,
where $G$ is defined in Eq.(\ref{gfn}).
Aside from momentum prefactors, we need to evaluate
\begin{equation}
\label{app1.1}
I_{\alpha,\beta }({\bf \kappa},t) = 
\Biggl \langle \xi ^{\alpha }(t)\xi ^{\beta }(t) \exp \Bigl 
[i\int \limits _{0}^{t} dt' \ 
{\bf \xi}(t')\cdot {\bf \kappa}(t') \Bigr ] \Biggr\rangle \ .
\end{equation}
We have generalized the momentum in the exponent to a time-dependent
form ${\bf \kappa }(t)$ for a reason soon to become clear. At the
end of the averaging procedure we shall reset ${\bf \kappa }={\bf k}-
{\bf k}_{1}$ as required.

The average given above is most easily evaluated by generating the
noise prefactors via functional differentiation with respect to
${\bf \kappa}(t)$. Thus,
\begin{eqnarray}
\label{app1.2}
\nonumber
I_{\alpha,\beta }({\bf \kappa},t) & = & -{\delta ^{2} \over 
\delta \kappa ^{\alpha }(t)\delta \kappa ^{\beta }(t)}
\Biggl \langle \exp \Bigl [ i\int \limits _{0}^{t} dt' \ 
{\bf \xi}(t')\cdot {\bf \kappa}(t') \Bigr ] \Biggr \rangle\\ 
\nonumber
& = & -{\delta ^{2} \over 
\delta \kappa ^{\alpha }(t)\delta \kappa ^{\beta }(t)}
\exp \Bigl [ -(D/2)\int \limits _{0}^{t} dt' \ \kappa (t')^{2} \Bigr ] \\
& = & {D\over t_{0}} \left [ \delta _{\alpha,\beta }
- Dt_{0}\kappa ^{\alpha}(t) \kappa ^{\beta} (t) \right ]
\exp \Bigl [ -(D/2)\int \limits _{0}^{t} dt' \ \kappa (t')^{2} \Bigr ] \ ,
\end{eqnarray}
where $t_{0}$ is the implicit scale of the noise correlations.
We need only retain the first term, given the smallness of $t_{0}$.
Using this result with Eq.(\ref{mft1}) yields Eq.(\ref{mft2}) in
the main text.

\section{}
In this appendix we outline the evaluation of $Q_{n}$ as defined in
Eq.(\ref{qfn}). The terms in the exponent of $Q_{n}$ are of the
form ${\bf R}(t_{m})\cdot ({\bf k}_{m-1}-{\bf k}_{m})
= \int \limits _{0}^{t_{m}} ds_{m} \ 
{\bf \xi}(s_{m})\cdot ({\bf k}_{m-1}-{\bf k}_{m})$.
We replace the momenta appearing in the
exponent with generalized momenta ${\bf \kappa}_{m}(s_{m})$. After
the averaging procedure we set ${\bf \kappa}_{m}(s_{m}) = 
{\bf k}_{m-1}-{\bf k}_{m}$.
This allows us to generate the noise prefactors from functional 
differentiation. Thus we have
\begin{equation}
\label{app2.1}
Q_{n}(\lbrace {\bf \kappa}_{m},t_{m};\alpha _{m},\beta _{m}\rbrace ) = 
\left [ \prod \limits _{m=1}^{n} {\delta ^{2} \over 
\delta \kappa _{m}^{\alpha _{m}}(t_{m})
\delta \kappa _{m}^{\beta _{m}}(t_{m})}\right ] \ 
\left \langle 
\exp \left [ i\sum \limits _{m=1}^{n} \int \limits _{0}^{t_{m}}
ds_{m} \ {\bf \xi}(s_{m}) 
\cdot {\bf \kappa }_{m}(s_{m}) \right ] \right \rangle \ .
\end{equation}
The average appearing above is easily performed over the multivariate
gaussian noise distribution, and yields
\begin{equation}
\label{app2.2}
\left \langle 
\exp \left [ i\sum \limits _{m=1}^{n} \int \limits _{0}^{t_{m}}
ds_{m} \ {\bf \xi}(s_{m}) 
\cdot {\bf \kappa }_{m}(s_{m}) \right ] \right \rangle
= \exp \left \lbrace -{D \over 2} \sum \limits _{m=1}^{n} \int \limits
_{t_{m+1}}^{t_{m}} ds \ \left [\sum \limits _{l=1}^{m} 
{\bf \kappa}_{l}(s) \right ]^{2} \right \rbrace \ .
\end{equation}
A given double functional derivative of the exponent gives:
\begin{eqnarray}
\label{app2.3}
\nonumber
{\delta ^{2} \over 
\delta \kappa _{m}^{\alpha _{m}}(t_{m})
\delta \kappa _{m}^{\beta _{m}}(t_{m})} \exp \Bigl \lbrace -(D/2)
\int \limits _{t_{m+1}}^{t_{m}} ds \ [ {\bf K}(s)
& + & {\bf \kappa}_{m}(s)]^{2}
\Bigr \rbrace \\
\nonumber
= -{D\over t_{0}}\Bigl [ \delta _{\alpha _{m},\beta _{m} }
- Dt_{0}(K^{\alpha _{m}}(t_{m}) & + & \kappa ^{\alpha _{m}}(t_{m}))
(K^{\beta _{m}}(t_{m})+\kappa ^{\beta _{m}} (t_{m})) \Bigr ] \\
& \times & \exp \Bigl \lbrace -(D/2) \int
\limits _{t_{m+1}}^{t_{m}} ds \ [ {\bf K}(s) + {\bf \kappa}_{m}(s)]^{2}
\Bigr \rbrace \ .
\end{eqnarray}
We use this last result to perform the $n$ double functional derivatives
in Eq.(\ref{app2.1}). We note that using our final replacement
for $\lbrace {\bf \kappa}_{m} \rbrace$ yields $\sum \limits _{l=1}^{m}
{\bf \kappa}_{l} = {\bf k}-{\bf k}_{m}$. Thus we reproduce Eq.(\ref{qfn2})
as given in the main text.

\section{}
In this appendix we give a brief analysis of the integral appearing
in Eq.(\ref{ftdendlt2}), specializing to $d=1$. 
In terms of the density difference, this has the form
\begin{equation}
\label{app3.1}
\delta {\tilde \rho }({\bf k},t) = 
B{i \over 2\pi} \int _{C} d\tau \ {e^{\tau} \over \tau ^{1/2}}
\left [ {1 \over \tau ^{1/2} + b(k,t)}
\right ] \ ,
\end{equation}
where for convenience we have set $b=(k/\Lambda )^{2}(t/t_{0})^{1/2}$.
First, we change variables from $\tau $ to $-x$ (which runs along the
negative real axis) by integrating across the branch cut. This yields
\begin{equation}
\label{app3.2}
\delta {\tilde \rho }({\bf k},t) = 
{B \over \pi} \int \limits _{0}^{\infty} {dx \over \sqrt{x}} \ 
{e^{-b^{2}x} \over (1+x)} \ .
\end{equation}
Next we change variables to $y=\sqrt{x}$. Rewriting the exponential
using a Hubbard-Stratonovich transformation gives the double integral
\begin{equation}
\label{app3.3}
\delta {\tilde \rho }({\bf k},t) = 
{B \over \pi} \int \limits _{-\infty }^{\infty} {dy \over (1+y^{2})} \ 
\int \limits _{-\infty }^{\infty } {ds \over \sqrt{\pi }}
\ e^{-s^{2} + 2ibsy} \ .
\end{equation}
The integral over $y$ now resembles a Fourier transform, and results in a 
simple exponential, so that
\begin{equation}
\label{app3.4}
\delta {\tilde \rho }({\bf k},t) = 
B \int \limits _{-\infty }^{\infty} {ds \over \sqrt{\pi }}
\ e^{-s^{2} -2b|s|} \ .
\end{equation}
The inverse Fourier transform from $k$ to $r$ is now easily performed
resulting in Eq.(\ref{scalfn1d}), as shown in the main text.

For general $d \in [0,2]$, the analysis is more difficult. Progress
is made by exponentiating the Lorentzian form in Eq.(\ref{ftdendlt2})
using an auxiliary integral, and then performing the asymptotic 
expansions on the resulting double integral.


\newpage

\begin{references}

\bibitem{hug} B. D. Hughes, {\it Random Walks and Random Environments},
(Clarendon Press, Oxford, 1995); C. Itzykson and J.-M. Drouffe, 
{\it Statistical Field Theory}, (Cambridge University Press, 1989).

\bibitem{vmdr} J. R. Manning, {\it Diffusion Kinetics for Atoms in Crystals},
(Van Nostrand, Princeton N.J., 1968).

\bibitem{vmd} C. A. Scholl, J. Phys. C, {\bf 14} 2723 (1981);
K. W. Kehr, R. Kutner and K. Binder, Phys. Rev. B, {\bf 23}
4931 (1981);
H. Van Beijeren et al., Phys. Rev. B {\bf 28} 5711 (1983),
and references therein; R. G. Palmer et al, Phys. Rev. Lett., 
{\bf 53} 958 (1984).

\bibitem{bh} M. J. A. M. Brummelhuis and H. J. Hilhorst, J. Stat. Phys.,
{\bf 53} 249 (1988).

\bibitem{toro} Z. Toroczkai, Int. J. Mod. Phys., {\bf B11} 3343 (1997);
R. K. P. Zia and Z. Toroczkai, in press (1998).

\bibitem{fe} H. E. Stanley, {\it Introduction to Phase Transitions
and Critical Phenomena}, (Oxford University Press, 1971).

\bibitem{wk} K. G. Wilson and J. Kogut, Phys. Rep., {\bf 12} 75 (1974). 

\bibitem{sch} Z. Toroczkai et al., Europhys. Lett., {\bf 40} 281 (1997).

\bibitem{as} {\it Handbook of Mathematical Functions} 10th Edition, 
M. Abramowitz and I. A. Stegun Eds. (Dover, NY, 1972).

\bibitem{cup} J. C. Phillips, {\it Physics of High-Temperature
Superconductors}, (Academic Press, San Diego, 1989).

\bibitem{pin} {\it Phenomenology and Applications of High-Temperature
Superconductors}, eds. K. Bedell et al., (Addison-Wesley, New York, 1992).

\bibitem{ld} W. E. Lawrence and S. Doniach, in {\it Proceedings of the
12th International Conference on Low Temperature Physics}, ed. E. Kanda 
(Acad. Press of Japan, Kyoto, 1971).

\end{references}
\end{document}